**ON THE EXPANSION OF NON-IDEAL COPPER PLASMA INTO VACUUM**

Dmitry Levko, Robert R. Arslanbekov and Vladimir I. Kolobov

CFD Research Corporation, 701 McMillian Way, Huntsville, AL 35806, USA

The expansion of non-ideal copper plasma into vacuum is analyzed for the conditions typical to explosive electron emission in vacuum arcs. The gas-dynamic model solves the Euler equations with an equation of state (EoS) for weakly non-ideal plasma taking into account the ionization energy correction due to electron-ion Coulomb coupling. We have obtained that the EoS has insignificant influence on the plasma expansion, when the plasma properties and composition change drastically. Based on our simulation results, the validity of the "frozen" state theory often used in the vacuum arcs plasma diagnostics is questionable.

## I. Introduction

Plasmas expanding into vacuum or a background gas appear in many technological applications such as laser ablation,[1,2] physical vapor deposition,[3,4] electric propulsion,[5] inertial fusion,[6] hypersonic aircraft flow control, plasma assisted combustion,[7] *etc*.

Vacuum arc is one of the phenomena where plasma expansion appears naturally.[8] Today, it is argued that the ignition and operation of vacuum arcs occurs through a series of micro-explosions at the cathode surface.[8,9,10,11,12,13] These micro-explosions provide the gaseous media for a plasma plume forming the arc channel. The micro-explosion mechanism can be explained as follows.[9,10,12] Micro-protrusions, which are always present at the metal surface, cause local enhancement of the electric field. When the electric field at the micro-protrusion tip exceeds the critical value ($\sim 2 \times 10^7$ V/cm), the electron field emission from the tip starts. The electric current heats the micro-protrusion causing thermo-field emission. If the emission current is large enough ($> 2 \times 10^8$ A/cm$^2$ for copper cathode), a thermal runaway[9,14] occurs: the micro-protrusion temperature reaches the liquid metal boiling temperature on the nanosecond time scale as a continuum phase transition metal-liquid-gas-plasma.[8] On such a short time scale, the electrode material cannot move, and after a few nanoseconds, a dense plasma cloud appears at the cathode surface. Very similar processes take place during the femto-second laser ablation when a solid-to-plasma phase transition occurs on the sub-nanosecond time scale under the action of a laser pulse.[2]

The micro-explosions are accompanied by strong pressure gradients, which result in ion acceleration to supersonic velocities.[8,15,16,17] In Ref. 18, it was pointed out that measuring ion velocity distribution function far from the explosion center can give useful information about the plasma state in the explosion center. This method of plasma diagnostics is based on a theory of "frozen" state.[18] This theory assumes that the transition from dense equilibrium plasma to dilute non-equilibrium plasma occurs in an infinitely thin layer. In this layer, the plasma composition remains frozen and does not change during the plasma expansion.

The expansion of dense plasma from an explosive emission center was studied in numerous publications[19,20,21,22] using both hydrodynamic and particle-in-cell (PIC) models. Since the plasma density in the explosive emission center is close to solid density ($\sim 10^{28}\,\text{m}^{-3}$),[8] the use of PIC models for the description of explosive emission center is questionable because the traditional PIC methods fail at high plasma densities, when the number of electrons within the Debye sphere exceeds the unity.[23] On the other hand, hydrodynamic models using ideal gas law for expanding plasmas miss important effects related to non-ideal character of dense plasmas. In particular, the non-ideal gas equation of state (EoS) and the pressure ionization must be taken into consideration.[8]

As was pointed out in Ref. 24, vacuum breakdown occurs on the ion time scale, while the gaseous breakdown occurs on the electron time scale.[25] The disparity of time scales (electron, ion and liquid) allows significant simplification of the computational models. The liquid and electron time scales were explored in our previous publication.[25] In the present paper, we explore the processes that occur on the ion time scale. Here, we analyze the expansion of weakly non-ideal plasma into vacuum using a fluid model with a non-ideal plasma EoS. This EoS takes into account the contributions from both neutrals and charged species, and the ionization energy correction due to electron-ion Coulomb coupling. We consider only the gas dynamic model based on Euler equations, and assume that the electrons and ions are in the local thermodynamic equilibrium (LTE). By varying the gas temperature and pressure in a wide range we analyze how the plasma non-ideality influences the plasma expansion dynamics.

## II.    Computational model

There are two regimes of plasma expansion,[18,26] which are characterized by the Damköhler number, $Da$ (see Figure 1), which is the ratio of the characteristic expansion time and the characteristic reaction time. At $Da \ll 1$, the plasma is in thermal non-equilibrium, while the ionization and recombination processes can be neglected. Then, the charge states are said to be frozen and the ion charge distribution function is not changing during the expansion.[18] In the opposite case, at $Da \gg 1$, the charge state distribution is in the LTE. This means there is a balance between the ionization and recombination reactions due to frequent interparticle collisions. In a more convenient form, the necessary condition for LTE is defined by the McWhirter criterion[27]

$$n_e(m^{-3}) = 1.6 \times 10^{18} T^{1/2} (\Delta E_{nm})^3, \tag{1}$$

where $T$ is the plasma temperature (in K) and $\Delta E_{nm}$ is the energy difference between levels $n$ and $m$ (in eV).

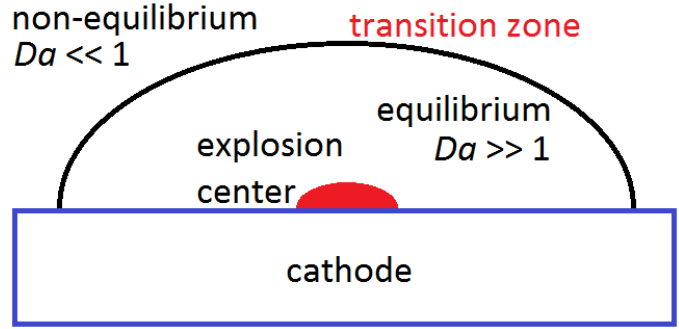

Figure 1. Schematic representation of plasma expansion.

Plasma is quasi-neutral in the LTE regime. As was shown, for instance, in Ref. 26, this regime is realized in the vicinity of the explosion center ($x < 1$ mm), where the gas density is close to the solid density. As the plasma expands, its density decreases. Then, the balance between ionization and recombination breaks, and the plasma comes to the non-thermal equilibrium state. In the present paper, only the LTE plasma expansion is considered. The expansion of non-equilibrium plasma was analyzed, for instance, in Ref. 28 using the kinetic Vlasov equation solver. There are also numerous theoretical and hydrodynamic modeling studies devoted to the expansion of this plasma (see Refs. 29,30,31 and references therein).

The LTE plasma expansion into vacuum can be described by the system of one-dimensional (1D) Euler equations:[1]

$$\frac{\partial \rho}{\partial t} + \frac{\partial (\rho u)}{\partial x} = 0, \tag{2}$$

$$\frac{\partial (\rho u)}{\partial t} + \frac{\partial}{\partial x}\left(p + \rho u^2\right) = 0, \tag{3}$$

$$\frac{\partial}{\partial t}\left[\rho\left(E + \frac{u^2}{2}\right)\right] + \frac{\partial}{\partial x}\left[\rho u\left(E + \frac{p}{\rho} + \frac{u^2}{2}\right)\right] = 0. \tag{4}$$

Here, $\rho$ is the density, $u$ is the velocity, $p$ is the pressure and $E$ is the internal energy. The system (2)-(4) is solved for the density $\rho$, the momentum $\rho u$, and the internal energy density $\rho E$. A multi-dimensional, ideal-gas version of this model has been recently developed in Ref. 32. This model relies on a different compressible Euler solver and uses two Cu ion species, $Cu^+$ and $Cu^{2+}$. The model is thus applicable mainly to low-density, ideal plasma conditions, when the ionization into higher Cu states is not important. The 1D version of this model has been successfully validated against laser ablation experiments.

The system (2)-(4) must be closed by the appropriate EoS. Dense plasma can be characterized by the degeneracy parameter, $\xi$, and the ideality parameter, $\gamma$.[33] The parameter $\xi$ is the ratio of the plasma temperature and the Fermi energy, while $\gamma$ is the ratio of the potential and kinetic energies of plasma particles. For the explosive emission plasma, one has $\xi_e < 1$ and $\xi_i < 1$, while $\gamma > 1$.[9] This means that such plasma can be described by the classical statistical theory, although there is a strong interaction between the electrons and ions.[33] Therefore, in our present model we assumed that each plasma species is described by the ideal gas EoS resulting in the total pressure

$$p = (1 + x_e)\rho kT/m, \tag{5}$$

while the internal energy density is defined by[1]

$$\rho E = \frac{\rho}{m}\left[\frac{3}{2}(1+x_e)kT + \sum_{N=1}^{Z_{max}}\left(E_N\sum_{Z=1}^{Z_{max}}x_z\right)\right].\tag{6}$$

Here, $m$ is the atom mass, $x_e$ is the electron fraction, $x_z$ is the ion fraction, and $Z_{max}$ is the total number of ions considered in the model. We also impose the conservation of mass[1]

$$x_0 + \sum_{Z=1}^{Z_{max}}x_z = 1,\tag{7}$$

and the charge conservation

$$\sum_{Z=1}^{Z_{max}}Zx_z = x_e.\tag{8}$$

In Eq. (7), $x_0$ is the atoms fraction. Within the LTE, the densities of plasma species are calculated using the system Saha-Eggert equations:[1,18]

$$\frac{n_e n_{Q+1}}{n_Q} = \Lambda_B^{-3}\frac{2\Sigma_{Q+1}}{\Sigma_Q}\exp\left(-\frac{E_Q - \Delta E_Q}{kT}\right).\tag{9}$$

Here, $\Lambda_B = \frac{h}{\sqrt{2\pi m_e kT}}$ is the thermal de Broglie wavelength, $h$ is the Planck constant, and $\Sigma_Q$ are the ion partition functions. Equation (9) takes into account both the thermal and pressure ionization.[34] The latter effect is obtained at high pressures, when the outer shells of atoms become compressed and finally disappear.[33] The ionization threshold lowering ($\Delta E_Q$) due to the electron-ion Coulomb coupling is calculated using the Debye-Hückel theory:[35]

$$\Delta E_Q = \frac{(Q+1)e^2}{4\pi\varepsilon_0(\lambda_D + \Lambda_B/8)}.\tag{10}$$

Here, $e$ is the elementary charge, and $\lambda_D$ is the plasma Debye length calculated by[35]

$$\lambda_D = \left\{\frac{\varepsilon_0 kT}{e^2(n_e + \sum_Q Q^2 n_Q)}\right\}^{\frac{1}{2}}.\tag{11}$$

Here, $\varepsilon_0$ is the permittivity of a free space, $n_e$ is the electron number density, and $n_Q$ are the ion number densities. The Debye-Hückel theory is valid until $\Delta E_Q < E_Q$.[35]

The numerical solution of Eqs. (1)-(3) requires the calculation of sound speed.[36] The latter was found from

$$c_s^2 = \frac{\partial p}{\partial \rho}\Big|_E + \Gamma\frac{p}{\rho}, \tag{12}$$

where $\Gamma = \frac{1}{\rho}\frac{\partial p}{\partial e}\Big|_\rho$ is the Grüneisen coefficient.[37]

### III.    Global analysis of plasma composition

This section presents the global analysis of the plasma composition. In this analysis, only the system of equations (7)-(11) was solved. The temperature was considered as the external parameter.  The copper (Cu) plasma consisting of 29 ions in the ground states was considered. The partition functions and the ionization energies were taken from Refs. 18 and 38, respectively. The gas density was varied from the rarefied flow conditions ($10^{-6}$ kg/m$^3$) (Knudsen number ~1-10) to the solid state ($10^4$ kg/m$^3$) conditions.

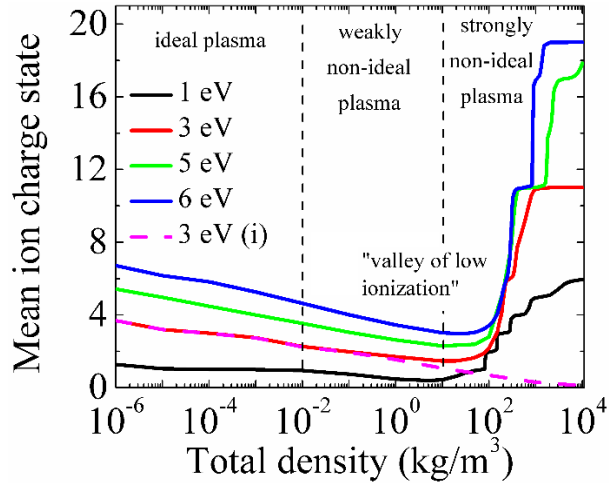

Figure 2. Influence of the gas density on plasma composition for different values of the temperature. Dash line shows the results obtained for 3 eV without accounting for the lowering of ionization energy.

Figure 2 shows the influence of the total density and the temperature on the mean ion charge state, $x_m$. Similar results for various metals were obtained, for instance, in Refs. 18,39,40,41.  In our model, $x_m = x_e$ because the total species density is kept constant.

The pink dash line in Figure 2 shows the mean ion charge state obtained for 3 eV without accounting for the lowering of ionization energy. One can see that below $10^{-2}$ kg/m$^3$, this curve coincides with the one obtained with account for the ionization energy correction. This means that this region on the diagram corresponds to ideal plasma. In the region $10^{-2}$ kg/m$^3$< $\rho$ <$10^1$ kg/m$^3$, both models predict a decrease of $x_m$ for increasing density. However, the model with the ionization energy correction predicts a minimum of $x_m$ at $\rho$~ 20 kg/m$^3$, while another model predicts a monotonically decreasing $x_m$. In this model, only the thermal gas ionization is possible. Therefore, at extremely high gas density, there is not enough energy for the gas ionization. In Ref. 18, the region in the vicinity of the minimum of curve $x_m = x_m(\rho, T)$ was called the "valley of low ionization". In this region, the transition from weakly to strongly non-ideal plasma occurs.

The increase of total plasma density leads to an increase of the role of the electron-ion Coulomb coupling, which results in a sharp increase of the mean ion charge. In this region, the main mechanism of the gas ionization is the pressure ionization.[34] Figure 2 shows that at high densities, the value of $x_m$ saturates. One can conclude that the lower the temperature the lower the saturation value of $x_m$. In the asymptotic limit of very high pressures (not reachable for the conditions of our studies) and high temperatures, the plasma is fully ionized. It consists only of free, bare nuclei and free electrons.

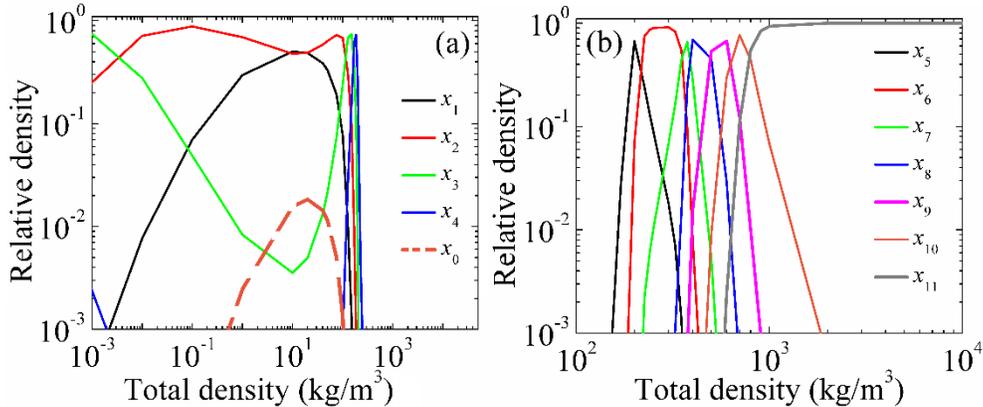

Figure 3. Plasma composition as function of gas density obtained for the temperature of 3 eV.

Figure 3 shows the plasma composition as function of total density for $T$ = 3 eV. These results were obtained using the model with accounting for the ionization energy lowering. One can see that there is a wide region where the ion with $Z = 2$ is the dominant. This region is typical for the laser ablation plasma.[1] In the region $10^2$ kg/m$^3$ < $\rho$ < $10^3$ kg/m$^3$, there are several ions with the charge state $5 < Z < 11$ present in the plasma. However, there is always one dominant ion, and the densities of other ions are much smaller. At $\rho > 2 \times 10^3$ kg/m$^3$, only the ions with $Z = 11$ are present in the plasma. This result is explained by the big energy gap between the energy levels #11 and #12 (~100 eV).[38]

## IV.    Analysis of the plasma expansion

In this section, we analyze the plasma expansion in the vicinity of the valley of low ionization (see Figure 2), where the Debye-Hückel theory is still valid. The gas stair with the initial pressure of $p_0$ and density $\rho_0$ was located in the left part of the simulation domain. On the right, the "vacuum" conditions were imposed (pressure $10^{-5}$ Pa and density $10^{-3}$ kg/m$^3$). Below, the results of simulations for $p_0 = 5 \times 10^9$ Pa and $\rho_0 = 10^3$ kg/m$^3$ are presented. These initial conditions are close to the ones obtained after the micro-protrusion explosion.[9] Here, we do not consider the fast solid-liquid-gas-plasma continuum phase transition. As was shown in Refs. 9,14, this transition occurs at the nanosecond time scale due to the thermal runaway. This time is too short for the liquid motion, and the dynamics of this phase transition can be modeled by solving only the thermal heat balance equation.[9,14]

In the regime considered here, the pressure ionization is important, and one needs to consider the ionization energy correction defined by Eq. (10). For higher densities, plasma becomes strongly non-ideal, and the EoS (5)-(6) loses its validity. As it follows from Figure 2, for $T < 6$ eV and for the total density in the range $10^{-4} - 10^3$ kg/m$^3$, the mean ion charge state does not exceed 6. Therefore, in this section we consider the Cu plasma consisting only of 6 ions.

Figure 4-Figure 6 illustrate the main results of our simulations. These figures show a comparison between the models with and without accounting for the ionization potential lowering. One can conclude from Figure 4 that both models predict very similar results. Both the gas density and the velocity profiles look similar. The comparison between Figure 4(b) and Figure 4(d) shows that the ideal plasma propagates faster that the non-ideal one. This difference can be explained by the small changes in the gas pressure obtained for both plasmas.

One can conclude from Figure 4 and Figure 5 that both the velocity and temperature increase at the leading edge of the expanding plasma. Since the system (2)-(4) is isentropic,[42] the initial internal energy is gradually transformed into kinetic energy. The velocity increase follows from the density continuity equation (2), which shows that the decrease of density leads to the increase of the gas velocity.

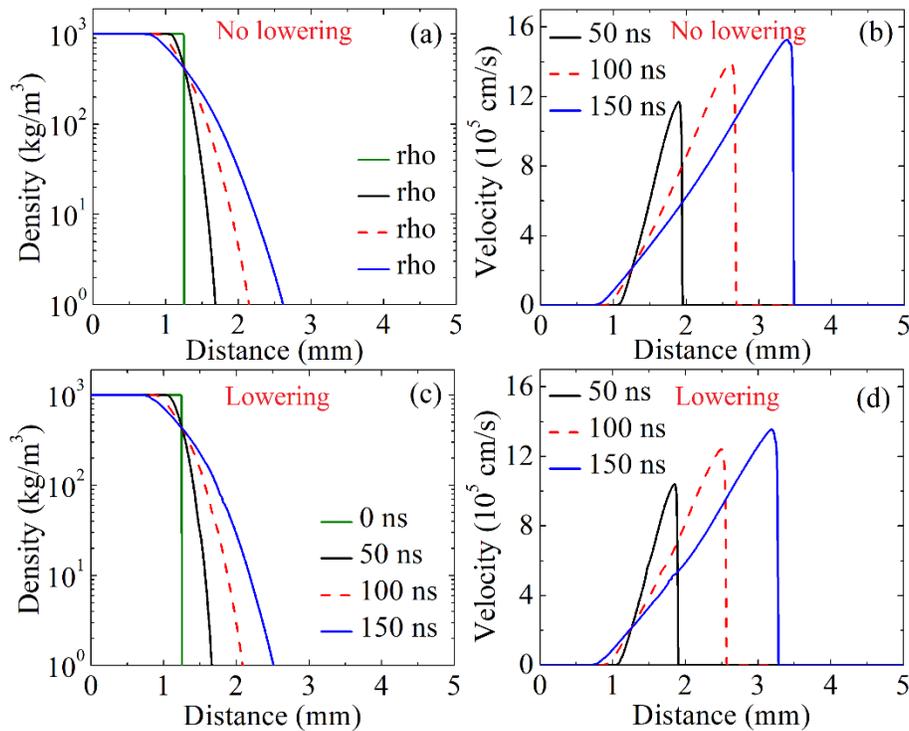

Figure 4.(a,c) Density and (b,d) velocity without and with the accounting for the ionization potential lowering for the plume pressure $5 \times 10^9$ Pa and density $10^3$ kg/m$^3$.

Although the ionization potential lowering does not influence significantly the plasma expansion (Figure 4), it does influence the plasma component content. Figure 5 shows the plasma temperature and the relative density of electrons, while Figure 6 shows the densities of main ions. One can see that the accounting for the pressure ionization results in the decrease of the temperature of the explosive emission center, while the electron relative density increases. Figure 6(d) shows that $x_e \sim 1$ in the plume of weakly non-ideal plasma, while it is ~0.3 in the plume of ideal plasma. This relation between $T$ and $x_e$ is obtained due to the way we set up the initial conditions. In our model, we fixed the initial pressure and density in the explosion center. Therefore, as it follows from the EoS (5), the value of $(1 + x_e)T$ does not depend on the model. The lowering of the ionization potentials in the Saha-Eggert equations (9) results in the higher densities of plasma species. This results in the decrease of the gas temperature calculated by Eq. (6).

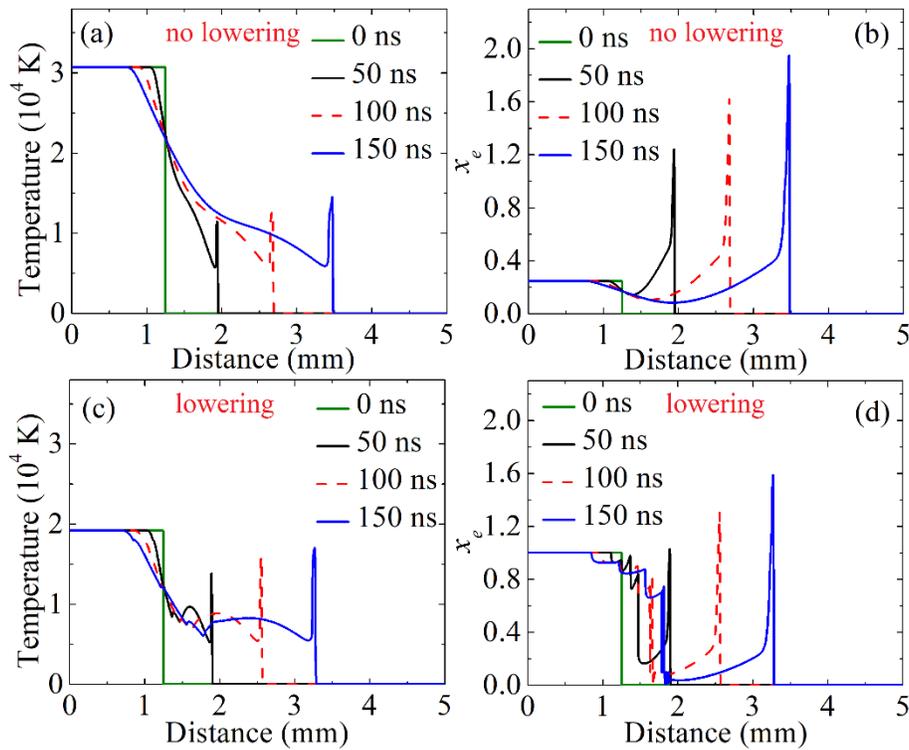

Figure 5. (a,c) Temperature and (b,d) electron relative density without and with the accounting for the ionization potential lowering for the plume pressure 5×10$^9$ Pa and density 10$^3$ kg/m$^3$.

Both models give comparable values of both temperature and $x_e$ at the leading edge of the expanding plasma (Figure 5). In this region, both the gas pressure and the temperature are smaller than in the explosion center. The pressure drops to such extent that the plasma becomes ideal at the leading edge.

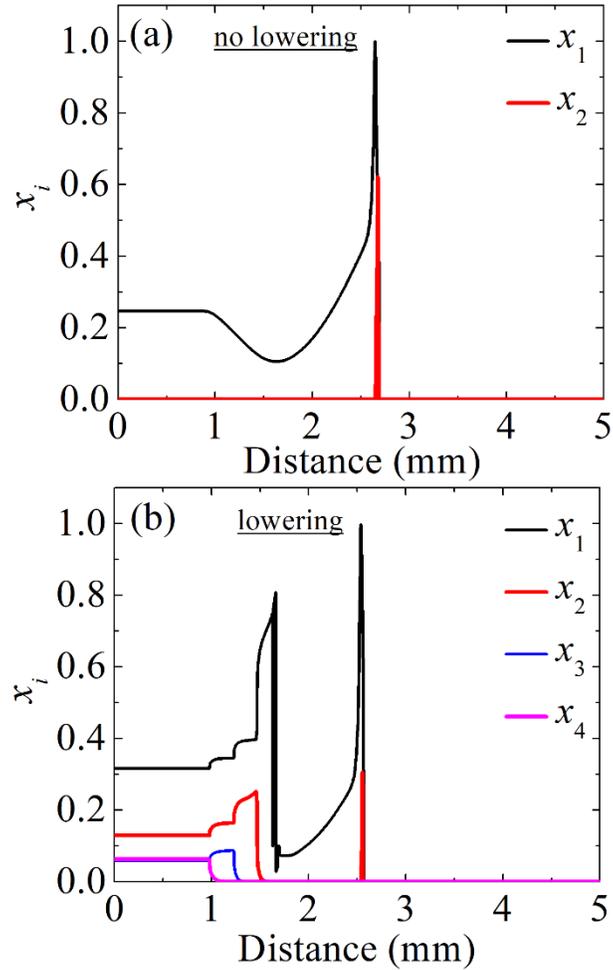

Figure 6. Primary ion densities obtained at $t = 100$ ns (a) without and (b) with the accounting for the ionization potential lowering for the plume pressure $5 \times 10^9$ Pa and density $10^3$ kg/m$^3$.

Figure 6 shows that there are only ions Cu$^+$ in the explosion center in the ideal plasma model. This is obtained because for the given gas density and temperature there is no enough energy for the thermal ionization with the generation of highly charged Cu ions. The accounting for the plasma non-ideality results in the appearance of the additional mechanism of the gas ionization, pressure ionization. This lowers the ionization energy thresholds in the Saha-Eggert equations (10) and results in the production of highly charged ions. Figure 6(b) shows that the non-ideal plasma of explosion center consists of four ions, Cu$^+$, Cu$^{2+}$, Cu$^{3+}$ and Cu$^{4+}$. Also, there is some small fraction of $Cu^{5+}$ ions in the explosion center ($x_5 \sim 10^{-4}$). One can see that the primary ions are $Cu^+$. There is also significant number of $Cu^{2+}$.

The simulation results show that outside the explosion center, the gas pressure decreases. This reduces the electron-ion Coulomb coupling and the plasma becomes ideal. Since the pressure ionization is switched off, only singly charged Cu atoms are obtained between the explosion center and the leading edge. The densities of highly charged ions drop almost to zero along the distance ~40 μm. Here, it is interesting to check the McWhirter criterion (1). For the conditions of Figure 5, one obtains that for $x < 2.5$ mm, the critical electron density is ~$10^{23}$ m$^{-3}$, while for $x > 2.5$ mm it is ~$10^{21}$ m$^{-3}$ (see Figure 7). One can then conclude that although the plasma species remain in the LTE at the distance ~1 mm from the explosion center, the plasma composition changes drastically. This is explained by the switching off of the pressure ionization which is obtained due to the plasma expansion. *This result does not agree with the theory of a frozen state,*[18] which neglects the transition from the pressure ionization dominant regime to the thermal ionization dominant regime along the finite distance. As it follows from our simulations, the plasma remains in the LTE during this transition. One possible explanation of this contradiction can be the inaccuracy of the McWhirter criterion (1) derived for the steady-state conditions.[27] As was discussed in Ref. 43, the accounting for the plasma propagation can increase the plasma density in Eq. (1) by at least a factor of 10.

Figure 5 shows that at the leading edge, the gas temperature increases. This leads to the generation of doubly charged ions due to the thermal ionization mechanism [see Figure 6(b)]. Here, it is important to note that in the vicinity of the leading edge, the McWhirter criterion (1) is not satisfied. Here, the electrostatic effects might be important.[29]

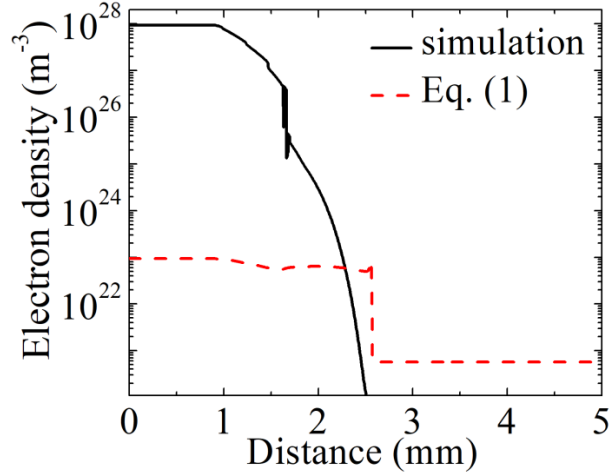

Figure 7. The McWhirter criterion (1) and the electron number density obtained at $t$ = 100 ns.

We also carried out the simulations for the explosion center pressure $5 \times 10^8$ Pa and density $10^2$ kg/m³. The obtained results were similar with the results shown above. The only difference was in the plasma composition. Namely, the decrease of the gas pressure results in the decrease of the energy correction factors $\Delta E_Q$. Therefore, the mean ion charge state decreases when both $p_0$ and $\rho_0$ decrease. For further decrease of the gas density, as it follows from Figure 2, the mechanism of thermal ionization becomes dominant, while the pressure ionization becomes less important. Our simulation results have also shown that the mechanism of plasma expansion qualitatively remains similar with the one presented in this section for higher gas pressure and density.

## V. Summary

The gas dynamic expansion of weakly non-ideal plasma into vacuum was analyzed for the conditions typical to explosive electron emission. For our analysis, we have developed the computational model which solves the Euler equations for gas dynamics taking into account the equation of state for non-ideal plasma with the ionization energy correction due to the electron-ion Coulomb coupling.

We have obtained that the Coulomb coupling has insignificant influence on the plasma expansion. However, this effect is significant for the plasma composition in the explosive emission center. We have also obtained that the leading-edge plasma is ideal and the ionization energy lowering is not important.

Our simulation results have shown that the plasma remains in the local thermodynamic equilibrium at distances about 1 mm from the explosive emission center. However, the plasma expansion at this distance is accompanied by the transition from the pressure-ionization- dominant regime to the thermal-ionization-dominant regime, which leads to drastic changes of the ion composition within the LTE region. Our results make questionable the validity of the "frozen" state theory often used in vacuum arcs plasma diagnostics.

**ACKNOWLEDGMENTS**

This work is supported by the DOE SBIR project DE-SC0015746 and by the NSF EPSCoR project OIA-1655280.